\documentclass[a4paper]{jpconf}
\usepackage{graphicx}
\usepackage{iopams}
\begin{document}
\title{Probing pairing correlations in Sn isotopes using Richardson-Gaudin integrability}

\author{S~De~Baerdemacker\textsuperscript{1,2}, V~Hellemans\textsuperscript{3}, R~van~den~Berg\textsuperscript{4}, J-S Caux\textsuperscript{4}, K~Heyde\textsuperscript{2}, M~Van~Raemdonck\textsuperscript{1,2}, D~Van~Neck\textsuperscript{1,2}, P~A~Johnson\textsuperscript{5}
}

\address{\textsuperscript{1} Ghent University, Center for Molecular Modeling, Technologiepark 903, 9052 Ghent, Belgium\\
\textsuperscript{2} Ghent University, Department of Physics and Astronomy, Proeftuinstraat 86, 9000 Ghent, Belgium\\
\textsuperscript{3} Universit\'e Libre de Bruxelles, PNTPM, CP229, 1050 Brussels, Belgium\\
\textsuperscript{4} Institute for Theoretical Physics, University of Amsterdam, Science Park 904, Postbus 94485, 1090 GL Amsterdam, The Netherlands\\
\textsuperscript{5}Department of Chemistry and Chemical Biology, McMaster University, Hamilton, Ontario, Canada
}

\ead{stijn.debaerdemacker@ugent.be}

\begin{abstract}
Pairing correlations in the even-even $A=102-130$ Sn isotopes are discussed, based on the Richardson-Gaudin variables in an exact Woods-Saxon plus reduced BCS pairing framework.  The integrability of the model sheds light on the pairing correlations, in particular on the previously reported sub-shell structure.
\end{abstract}

\section{Introduction}

Pairing is an important component of the correlations in atomic nuclei at low-excitation energy \cite{heyde:1994,dean:2003,brink:2005}.  The Sn isotopes provide a unique laboratory to probe the neutron-neutron pairing correlations, because the large proton shell gap at $Z=50$ ensures that the low-lying nuclear structure is largely unaffected by proton particle-hole excitations across the shell gap.  Moreover, experimental data of the Sn isotopes in three major shells have become available in recent years thanks to intensive experimental activity with radio-active beam facilities.  There exist several theoretical approaches to investigate pairing correlations in atomic nuclei, ranging from fundamental ab initio calculations to studies based on a more phenomenological footing \cite{dean:2003}.  In the present contribution, we will employ a  Woods-Saxon \cite{bohr:1998} plus level-independent Bardeen-Cooper-Schrieffer (BCS) pairing Hamiltonian \cite{bardeen:1957,bohr:1958} as a global probe for pairing correlations in the ground state of Sn.  The level-independent, or reduced, BCS Hamiltonian has a complete basis of Bethe Ansatz eigenstates \cite{richardson:1963,richardson:1964a}, and belongs to the class of Richardson-Gaudin (RG) integrable models \cite{gaudin:1976,dukelsky:2004a}.  Integrability offers unique opportunities to investigate pairing correlations.  On the one hand, the RG variables in the pair-product structure allow for a transparent graphical representation, as well as a clear-cut connection with bosonization approximations \cite{ring:2004} via a pseudo-deformation of the quasi spin algebra \cite{debaerdemacker:2012b}.  On the other hand, physical observables related to particle removal and addition properties \cite{grasso:2012} can be obtained conveniently using Slavnov's theorem for the RG model \cite{faribault:2008}.  

\section{Richardson-Gaudin integrability for Sn isotopes }

\begin{figure}[!htb]
\begin{center}
 \includegraphics{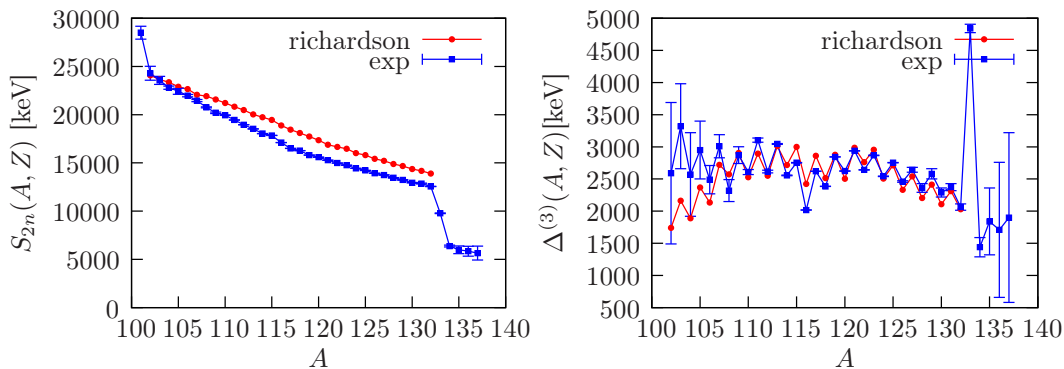}
 \caption{Experimental (squares) and theoretical (circles) two-neutron separation energies $S_{2n}$ (a) and three-point pairing gaps $\Delta^{(3)}$ (b).  Experimental data taken from \cite{audi:2003}.}\label{figure:S2nDelta}
\end{center}
\end{figure}

\begin{figure}[!htb]
 \begin{center}
 \includegraphics{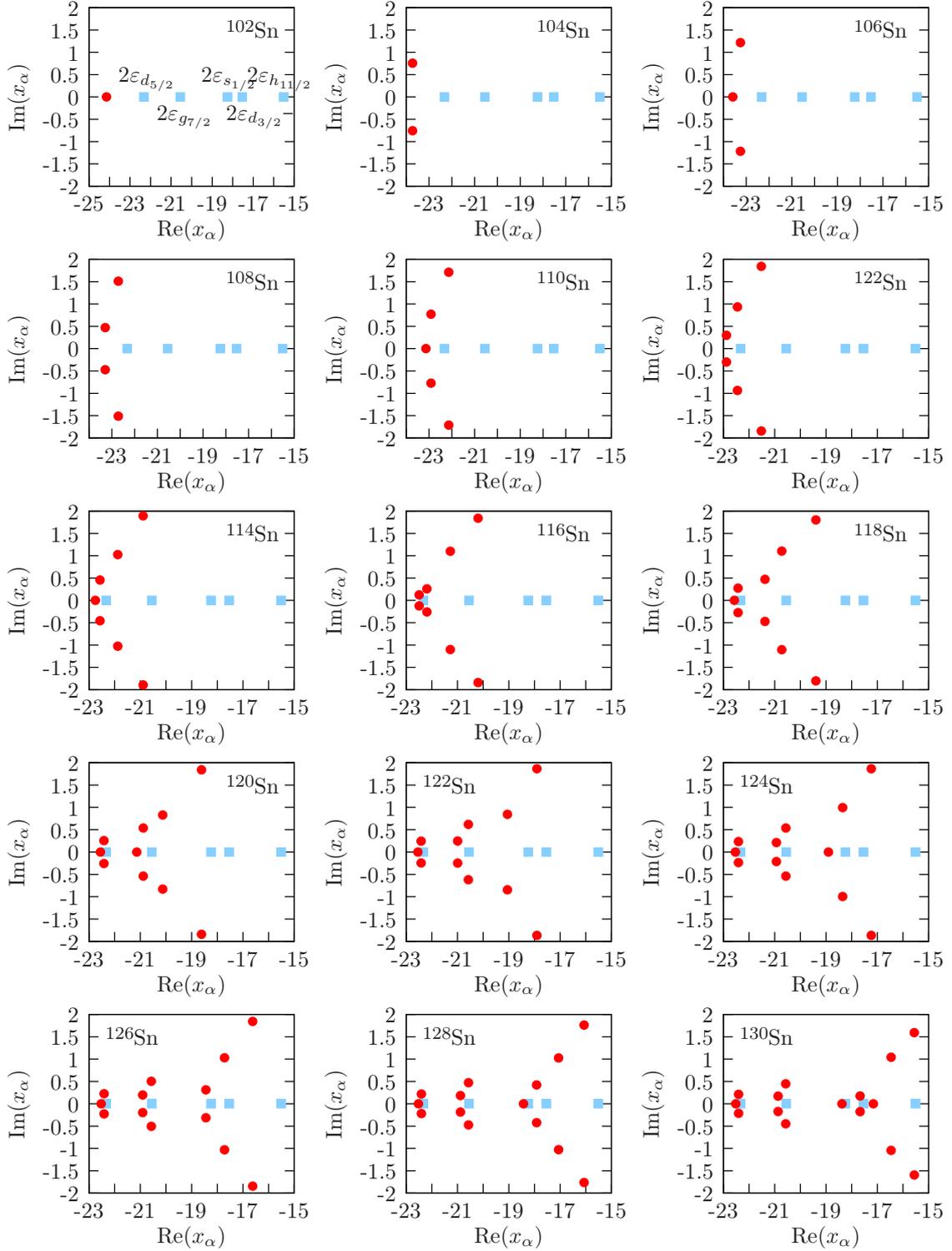}
 \caption{The RG variables (circles) and single-particle poles (squares) of the even-even \textsuperscript{102-130}Sn isotopes.}\label{figure:variables}
 \end{center}
\end{figure}

The reduced BCS Hamiltonian is given by \cite{heyde:1994} 
\begin{equation}\label{richardson:hamiltonian}
\hat{H}=\sum_{i=1}^m \varepsilon_i\hat{n}_i + g\sum_{i,k=1}^m\hat{S}^\dag_i \hat{S}_k,
\end{equation}
with $\hat{S}^\dag_i=\sum_{m_i>0}(-)^{j_i-m_i}a^\dag_{j_im_i}a^\dag_{j_i-mi}$ the nucleon-pair creation operator in a single-particle level $\varepsilon_i$ with (spherical) quantum numbers ($i\equiv n_i,l_i,j_i$) and of degeneracy $\Omega_i=2j_i+1$.  This Hamiltonian supports a complete set of Bethe Ansatz eigenstates parametrised by the set of RG variables $\{x\}$ that are a solution of the RG equations \cite{richardson:1963,richardson:1964a}.  The associated eigenstate energy is then given as $E=\sum_{\alpha=1}^{N_p}x_\alpha+\sum_{i=1}^m\varepsilon_i v_i$, with $v_i$ the seniority \cite{talmi:1993}, and $N_p$ the number of pairs.
The single-particle levels are provided by a Woods-Saxon potential \cite{bohr:1998}, for which we used a recent global parametrisation \cite{schwierz:2007}, and the single-particle energy spectrum for \textsuperscript{100}Sn is given in Table \ref{table:tdadecomposition}.  We followed a global prescription $g=g_0/\sqrt{A}$ for the pairing interaction, in order to reproduce the 3-point pairing gaps $\Delta^{(3)}(A)=(-)^A[BE(A)-2BE(A-1)+BE(A-2)]$ \cite{bohr:1998}, presented in Figure \ref{figure:S2nDelta}b. 
The two-neutron separation energies $S_{2n}=[BE(A)-BE(A-2)]$ \cite{bohr:1998} are given in Figure \ref{figure:S2nDelta}a, following a general linear trend, with the exception of a small kink around mid shell, signaling a sub-shell closure.  The calculated curve is smoother than the experimental values at this point, consistent with the overestimated pairing gaps $\Delta^{(3)}$ around mid shell.  Recent measurements showed a decrease in the $B(E2:0^+_1\rightarrow 2^+_1)$ strength around mid shell \cite{jungclaus:2011}, which was qualitatively attributed \cite{morales:2011} to this sub-shell effect in the seniority scheme \cite{talmi:1993}.   Figure \ref{figure:variables} depicts the RG variables for the ground state of the even-even \textsuperscript{102-130}Sn isotopes, and sheds more light on the sub-shell structure.   Weakly correlated pair states give rise to a clustering of RG variables around the single-particle poles in the complex plane, whereas  collective pairing states organise the RG variables along a broad arc in the complex plane \cite{dukelsky:2004a,debaerdemacker:2012b}.  The pairing interaction in the lighter isotopes is strong enough to distribute the RG variables along an arc in the complex plane, however the arc only extends over the $d_{5/2}$ and $g_{7/2}$ sub-shell single particle poles.  For the heavier nuclei, the pairs separate into two distinct sets, with seven RG variables clustering around the $d_{5/2}$ and $g_{7/2}$ sub-shell poles and the remaining forming a collective arc around the other poles.  For medium-heavy nuclei, there is a gradual transition between both situations.  This structure can be quantified using the pseudo-deformation scheme, where all RG variables can be labeled according to their collective behaviour in the Tamm-Dancoff Approximation (TDA) (see Table \ref{table:tdadecomposition}) \cite{debaerdemacker:2012b}.  From the table, it can be seen that the TDA structure is consistent with the discussed sub-shell structure.  The lightest isotopes are consistent with a collective TDA condensation in the $d_{5/2}$ and $g_{7/2}$ sub shell, whereas the TDA structure of the heavier isotopes points towards a normal filling of the $d_{5/2}$ and $g_{7/2}$ sub shell, with the additional pairs collectively distributed over the  $s_{1/2}$, $d_{3/2}$ and $h_{11/2}$ sub shell.
\begin{table}[!htb]
\begin{center}
	\begin{tabular}{lr|ccccccccccccccc}
	level & $\varepsilon_i$ [MeV] & 1 & 2 & 3 & 4 & 5 & 6 & 7 & 8 & 9 & 10 & 11 & 12 & 13 & 14 & 15 \\
	\hline\hline
	$d_{5/2}$  & -11.164 & 1 & 2 & 3 & 4 & 5 & 6 & 7 & 8 & 3 & 3 & 3 & 3 & 3 & 3 & 3\\
	$g_{7/2}$  & -10.275 & 0 & 0 & 0 & 0 & 0 & 0 & 0 & 0 & 6 & 7 & 4 & 4 & 4 & 4 & 4\\
	$s_{1/2}$  &  -9.124 & 0 & 0 & 0 & 0 & 0 & 0 & 0 & 0 & 0 & 0 & 4 & 5 & 6 & 1 & 1\\
	$d_{3/2} $ &  -8.766 & 0 & 0 & 0 & 0 & 0 & 0 & 0 & 0 & 0 & 0 & 0 & 0 & 0 & 6 & 2\\
	$h_{11/2}$ &  -7.754 & 0 & 0 & 0 & 0 & 0 & 0 & 0 & 0 & 0 & 0 & 0 & 0 & 0 & 0 & 5\\
	\hline
	\end{tabular}
\end{center}
\caption{The single-particle energies $\varepsilon_{i}$ obtained from a Woods-Saxon potential \cite{schwierz:2007}, and the TDA eigenmode decomposition of the $0^+$ground state for the even-even isotopes \textsuperscript{102-130}Sn.  The number of active pairs $N_p$ in the isotope \textsuperscript{A}Sn is given in the upper row ($N_p=(A-50)/2$).}\label{table:tdadecomposition}
\end{table}
\section{Conclusions}
We have investigated pairing correlations in the Sn isotopes by inspecting the location of the RG variables with respect to the single-particle poles in the complex plane, generated by a schematic Woods-Saxon plus reduced BCS Hamiltonian.  The results point towards a sub-shell structure, consistent with previous studies.  We expect this structure to also be reflected in the relevant transition rates; this will be investigated in future publications
\ack
SDB is an FWO-Vlaanderen post-doctoral fellow and acknowledges an FWO travel grant for a "long stay abroad" at the University of Amsterdam (The Netherlands).  VH acknowledges financial support from the FRS-FNRS Belgium.  This project is also supported by Belspo IAP Grant no P7/12 (SDB, VH, and KH).

\section*{References}


\providecommand{\newblock}{}

\end{document}